\documentclass[multphys,vecphys]{svmult}

\usepackage{makeidx}         
\usepackage{graphicx}        
\usepackage{multicol}        
\usepackage[bottom]{footmisc}

\makeindex             

\newcommand{\msun}{\mbox{$M_{\odot}$}}

\newcommand{\Mcl}{\mbox{$M_{\rm c}$}}
\newcommand{\tdis}{\mbox{$t_{\rm dis}$}}

\newcommand{\Rn}{\mbox{$R_{\rm n}$}}

\newcommand{\rhon}{\mbox{$\rho_{\rm n}$}}
\newcommand{\de}{\mbox{${\rm \Delta} E/|E_0|$}}
\newcommand{\dm}{\mbox{${\rm \Delta} M/|M_0|$}}

\newcommand{\rhs}{\mbox{$r^2_{\rm h}$}}

\newcommand{\Mc}{\mbox{$M_{\rm c}$}}

\newcommand{\Mn}{\mbox{$M_{\rm n}$}}
\newcommand{\Mcloud}{\mbox{$M_{\rm n}$}}

\newcommand{\rh}{\mbox{$r_{\rm h}$}}

\newcommand{\vmax}{\mbox{$V_{\rm max}$}}

\newcommand{\ac}{\mbox{$a_{\rm c}$}}

\begin{document}

\title*{The effect of giant molecular clouds on star clusters}
\author{M. Gieles\inst{1,2}\and
S.F. Portegies Zwart\inst{2}\and
E. Athanassoula\inst{3}}
\institute{Utrecht University
\texttt{gieles@astro.uu.nl}
\and University of Amsterdam
\texttt{spz@science.uva.nl}
\and Observatoire de Marseille
\texttt{lia@oamp.fr}
}
%
%
\maketitle

%

\section{When a star cluster meets a cloud}
\label{sec:1}
\begin{figure*}
\centering
\includegraphics[height=3.8cm]{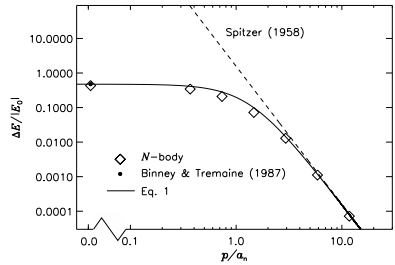}
\caption{\de\ of a cluster as a function of $p$. The $N$-body results
    are shown with diamonds. The result of \cite{1987gady.book.....B}
    and \cite{1958ApJ...127...17S} for head-on and distant encounters
    are shown as a filled circle and as a dashed line,
    respectively. Eq.~\ref{eq:1} is shown as a full line.}
\label{fig:1}      
\end{figure*}

We study the encounters between stars clusters and giant molecular
clouds (GMCs) \cite{gieles06b}. The effect of these encounters has
previously been studied analytically for two cases: 1) head-on
encounters, for which the cluster moves through the centre of the GMC
\cite{1987gady.book.....B} and 2) distant encounters, where the
encounter distance $p>3\Rn$, with $p$ the encounter parameter and \Rn\ the
radius of the GMC \cite{1958ApJ...127...17S}. We introduce an
expression for the energy gain of the cluster due to GMC encounters
valid for all values of $p$ and \Rn\ of the form

\begin{equation}
\Delta E  \simeq \frac{4.4\,\rhs}{\left(p^2+\sqrt{\rh\,\Rn^{3}}\right)^2}\,\left(\frac{G\Mcloud}{\vmax}\right)^2\,M_c.
\label{eq:1}
\end{equation}
Here \vmax\ is the maximum relative velocity, $\Mn$ is the mass of the
GMC, $G$ is the gravitational constant and \rh\ and \Mc\ are the
half-mass radius and mass of the cluster, respectively. We perform
$N$-body simulations of encounters with different $p$ and compare the
resulting ${\rm \Delta} E$ of the cluster to
Eq.~\ref{eq:1}. Fig.~\ref{fig:1} shows the very good agreement between
simulations and predictions of Eq.~1.  Snapshots of one simulation are
shown in Fig.~2.

\begin{figure*}
\centering
\includegraphics[height=2.35cm]{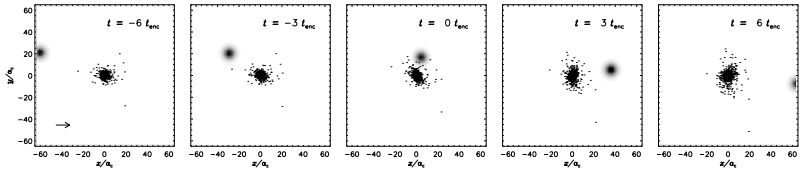}
\caption{Simulation of a close encounter between a GMC (grey shades)
and a star cluster. The snapshots are viewed in the centre-of-mass
frame of the cluster. \ac\ is the Plummer radius of the cluster.}
\label{fig:2}       
\end{figure*}
\vspace{-0.8cm}
\section{The cluster disruption time}
From the simulations we find that the fractional mass loss ($\dm$) is only
25\% of \de. This is because stars escape with velocities much higher
than the escape velocity. Defining the cluster disruption time as
$\tdis=\Mc/\dot{\Mc}$, we find a cluster disruption time of the form

\begin{equation}
\tdis = 2.0\,S\left(\Mc/10^4\,\msun\right)^{\gamma}{\mbox{Gyr}},
\label{eq:2}
\end{equation}
with $S\equiv1$ for the solar neighbourhood and scales with the global
 GMC density ($\rhon$) as $S\propto\rhon^{-1}$. The index $\gamma$ is
 defined as $\gamma=1-3\lambda$, with $\lambda$ the index that relates
 the cluster half-mass radius to its mass ($\rh \propto
 \Mcl^{\lambda}$).  The observed shallow relation between cluster
 radius and mass (e.g. $\lambda\simeq0.1$), makes the index
 ($\gamma=0.7$) similar to the index found both from observations
 \cite{2005A&A...441..117L} and from simulations of clusters
 dissolving in tidal fields ($\gamma\simeq0.62$). The constant of 2.0
 Gyr, which is the disruption time of a $10^4\,\msun$ cluster in the
 solar neighbourhood, is about a factor of 3.5 shorter than found from
 earlier simulations of clusters dissolving under the combined effect
 of the galactic tidal field and stellar evolution. It is only slightly
 higher than the observationally determined value of 1.3 Gyr
 \cite{2005A&A...441..117L}, suggesting that the combined effect of
 tidal field and encounters with GMCs can explain the lack of old open
 clusters in the solar neighbourhood \cite{1958RA......5..507O}. GMC
 encounters can also explain the (very) short disruption time that was
 observed for star clusters in the central region of M51
 \cite{2005A&A...441..949G}, since there $\rhon$ is an order of
 magnitude higher than in the solar neighbourhood.

%
%
\vspace{-0.3cm}
%
%

%
%

%

%







\printindex
\end{document}